\documentclass[ama,Numbered]{sn-jnl}

\usepackage{graphicx}%
\usepackage{multirow}%
\usepackage{amsmath,amssymb,amsfonts}%
\usepackage{amsthm}%
\usepackage{mathrsfs}%
\usepackage[title]{appendix}%
\usepackage{xcolor}%
\usepackage{textcomp}%
\usepackage{manyfoot}%
\usepackage{booktabs}%
\usepackage{algorithm}%
\usepackage{algorithmicx}%
\usepackage{algpseudocode}%
\usepackage{listings}%

\usepackage{subcaption}
\theoremstyle{thmstyleone}%
\theoremstyle{thmstyletwo}%

\theoremstyle{thmstylethree}%

\begin{document}

\title[Article Title]{Towards AI Lesion Tracking in PET/CT Imaging: A Siamese-based CNN Pipeline applied on PSMA PET/CT Scans }


\author*[1,2]{\fnm{Stefan P.} \sur{Hein}}\email{stefan.hein@tum.de}

\author[2]{\fnm{Manuel} \sur{Schultheiss}}

\author[3]{\fnm{Andrei} \sur{Gafita}}

\author[1]{\fnm{Raphael} \sur{Zaum}}

\author[2]{\fnm{Farid} \sur{Yagubbayli}}

\author[4]{\fnm{Robert} \sur{Tauber}}

\author[1]{\fnm{Isabel} \sur{Rauscher}}

\author[1]{\fnm{Matthias} \sur{Eiber}}

\author[2,5]{\fnm{Franz} \sur{Pfeiffer}}
\equalcont{These authors contributed equally to this work.}

\author[1]{\fnm{Wolfgang A.} \sur{Weber}}
\equalcont{These authors contributed equally to this work.}

\affil[1]{\orgdiv{Department of Nuclear Medicine}, \orgname{Technical University of Munich}, \orgaddress{\city{Munich}, \postcode{81675}, \country{Germany}}}

\affil[2]{\orgdiv{Chair of Biomedical Physics, Department of Physics}, \orgname{TUM School of Natural Sciences, Technical University of Munich}, \orgaddress{\city{Garching}, \postcode{85748}, \country{Germany}}}

\affil[3]{\orgdiv{Division of Nuclear Medicine and Molecular Imaging, The Russell H. Morgan Department of Radiology and Radiological Science}, \orgname{Johns Hopkins University School of Medicine}, \orgaddress{\city{Baltimore}, \postcode{21205}, \state{MD}, \country{USA}}}

\affil[4]{\orgdiv{Department of Urology}, \orgname{Technical University of Munich}, \orgaddress{\city{Munich}, \postcode{81675}, \country{Germany}}}

\affil[5]{\orgdiv{Munich Institute of Biomedical Engineering}, \orgname{Technical University of Munich}, \orgaddress{\city{Garching}, \postcode{85748}, \country{Germany}}}

\abstract{
	
Assessing tumor response to systemic therapies is one of the main applications of PET/CT. Routinely, only a small subset of index lesions out of multiple lesions is analyzed. However, this operator dependent selection may bias the results due to possible significant inter-metastatic heterogeneity of response to therapy. Automated, AI based approaches for lesion tracking hold promise in enabling the analysis of many more lesions and thus providing a better assessment of tumor response. This work introduces a Siamese CNN approach for lesion tracking between PET/CT scans.
	
Our approach is applied on the laborious task of tracking a high number of bone lesions in full-body baseline and follow-up [$^{68}$Ga]Ga- or [$^{18}$F]F-PSMA PET/CT scans after two cycles of [$^{177}$Lu]Lu-PSMA therapy of metastatic castration resistant prostate cancer patients. Data preparation includes lesion segmentation and affine registration. Our algorithm extracts suitable lesion patches and forwards them into a Siamese CNN trained to classify the lesion patch pairs as corresponding or non-corresponding lesions. Experiments have been performed with different input patch types and a Siamese network in 2D and 3D. The CNN model successfully learned to classify lesion assignments, reaching a lesion tracking accuracy of 83 \% in its best configuration with an AUC = 0.91. For remaining lesions the pipeline accomplished a re-identification rate of 89 \%. We proved that a CNN may facilitate the tracking of multiple lesions in PSMA PET/CT scans. Future clinical studies are necessary if this improves the prediction of the outcome of therapies.\\

}



\maketitle

\newpage
\section{Introduction}\label{sec1}

The American Cancer Society statistics of 2023 describing a general rise of cancer incidence \cite{Siegel2023} and cancer still representing one of the most frequent causes of death, show the high importance of targeted therapies and both specific and sensitive diagnostics. The reliable assessment of tumor response to systemic therapy is essential for drug development and patient management. Especially the combination of Positron Emission Tomography (PET) and Computed Tomography (CT) in nuclear medicine serves as a superior technique for sensitive lesion detection\cite{Weber2007} applicable for a high number of various cancer types (e.g. \cite{Kasamon2007},\cite{VandenAbbeele2008},\cite{Roedl2009},\cite{Rauscher2016}).

With prostate cancer being predicted to be the most common malignancy, surpassing breast cancer by 2030\cite{Quante2016}, the extensive research in the recent years has shown new successful ways for prostate cancer diagnostics and therapy, and their combination - theranostics. The success of prostate-specific membrane antigen (PSMA)-targeted radioligands labeled with positron- or $ \gamma $-emitting isotopes for imaging or with $ \beta $- or $ \alpha $-emitting isotopes for therapy has rapidly led to new standards in prostate cancer management.\cite{Weber2023} While PSMA-based PET/CT-imaging has already become the main diagnostic tool for response monitoring in prostate cancer cases in Europe \cite{Eiber2018}, in the USA the usage of [$^{68}$Ga]Ga-PSMA-11\cite{Eder2012} is a rather young method, approved by the American Food and Drug Administration (FDA) in December 2020\cite{Hennrich2021}. The same ligand type used for therapy as [$^{177}$Lu]Lu-PSMA-617 (Pluvicto$ ^{\text{TM}} $) has just recently been approved by the FDA in March 2022\cite{Hennrich2022} and the European Medicines Agency in December 2022\cite{EMAPluvicto} for metastatic castration resistant prostate cancer patients. 

\begin{figure}[h]
	\centering
	\includegraphics[width=1.0\textwidth]{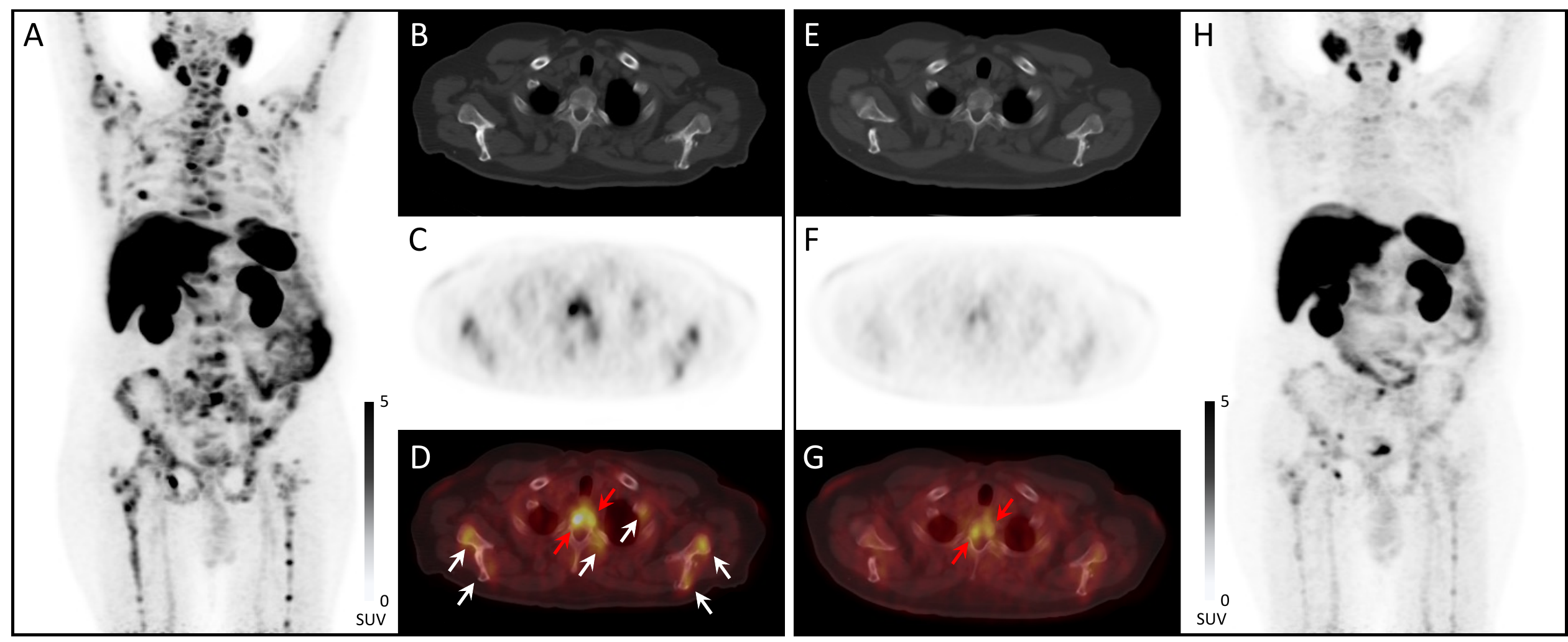}	
	\caption{\textbf{PSMA-PET/CT Scans showing a Male 73 year old Patient with Partial Response to [$^{177}$Lu]Lu-PSMA Therapy:} The baseline scan prior to therapy is shown on the left side, the follow-up scan after two therapy cycles on the right side: A\&H Maximum intensity projections (MIP), B\&E axial CT slices of the upper thorax, C\&F  corresponding axial PET slices, D\&G PET/CT overlay. The MIP show a clear reduction of tumor burdon with less bone lesions. Physiologic PSMA-uptake can be found in the lacrimal glands, salivatory glands, liver, spleen, small bowel, kidneys and bladder. The white arrows indicate resolved lesions, the red arrows remaining lesions. A complete comparative 3D analysis of all lesions is laborious and could be facilitated by an automated tracking algorithm.}
	\label{fig:Beispielbilder}
\end{figure}

The high sensitivity and specificity of PSMA-PET scans in detecting lesions together with the high anatomical contrast of the CT offers an enormous opportunity for accurate response monitoring and the detection of disease progression or regression in early stages. However, this creates new challenges to be solved. Especially prostate cancer patients frequently show diffuse bone lesion patterns with a high number of tracer uptake foci to be examined.\cite{Carlin2000} Each tumor response assessment requires the comparison of lesions in baseline and follow-up scan. Since detailed manual lesion tracking is laborious and error-prone, routinely, only a small subset of index lesions out of multiple lesions is analyzed.\cite{Wahl2009} However, the operator dependent selection of index lesions may bias the results due to possible significant inter-metastatic heterogeneity of response to therapy. Imaging with PSMA-radioligands might require a deeper and broader lesion analysis.

New automated and AI-based approaches for lesion tracking may enable the analysis of many more lesions providing a better assessment of tumor response. Also, the acquired data may offer the differentiation of therapy response for single body regions and show the impact of lesions in specific regions on the outcome of the disease. With the possibly highest number of PSMA-uptake foci and consequently the highest difficulty, automated lesions tracking in PET/CT scans seems to find its best use case for bone lesions of prostate cancer patients. For this reason, we focus on prostate cancer bone lesions for our presented new tracking approach.

So far, only few approaches exist for automated lesion tracking specifically in PET/CT scans. As the anatomy surrounding lesions is mostly similar, the techniques rely on image registration.\cite{Opfer2008}\cite{Fox2011} To our best knowledge, there is yet no tracking algorithm for PET/CT applying artificial intelligence, whereas several deep learning approaches are applied for segmentation or localization of lesions in PET scans\cite{Seifert2020}\cite{Sibille2020}.
Opfer et al. were among the first to suggest a tracking algorithm for lesions in PET/CT scans. It is based on a global rigid registration of consecutive CT scans and applies block matching and SUV region growing after a click on a selected baseline lesion.\cite{Opfer2008}
Fox et al. also proposes comparative analysis approach including a registration-based propagation of a selected region into the follow-up PET/CT scan, which has to be accepted or rejected by the reader.\cite{Fox2011}

In contrast to PET/CT scans, lesion tracking has been more studied for other imaging modalities, as ultrasound or stand-alone CT. There, registration-based approaches are common as well. Tan et al. use a deformable multiresolution B-Spline registration for evaluating treatment response in CT images of ovarial cancer patients.\cite{Tan2016} Within ultrasound and CT images, however, CNNs (Convolutional Neural Networks) have already been applied for lesion tracking. In Dankerl et al.\cite{Dankerl2014} a pattern recognition network detects landmarks in the follow-up scan and localizes follow-up lesions based on a patient specific graphical network. Furthermore, on the basis of a three-step-registration algorithm, Hering et al. use a nnU-Net for baseline CT lesion segmentation and perform lesion tracking by segmenting lesions in the propagated ROI of the follow-up CT scan.\cite{Hering2021}

In case deep learning is directly applied for lesion tracking, it is mostly in the shape of Siamese networks. In ultrasound images, they are applied by Gomariz et al.\cite{Gomariz2019}  and Liu et al.\cite{Liu2020} for  liver landmark and liver respiratory motion tracking. In CT scans, Rafael-Palou et al.\cite{Rafael-Palou2021} uses a 3D Siamese network for lung nodule tracking and Cai et al.\cite{Cai2021} extend a 3D Siamese network in depth for CT lesion tracking.

In this work, we aim to fill the gap of an AI lesion tracker for PET/CT scans, by applying a Siamese CNN structure in two and three dimensions on bone lesions in PSMA PET/CT images of metastatic castration resistant prostate cancer patients. With this, we select one of the most difficult lesion tracking applications.

We created a new fully-automated algorithm, tracking all existing lesions in baseline and follow-up scan by assigning corresponding lesions as well as recognizing resolved, new or fused lesions. After applying lesion segmentation and whole-body affine registration, a case distinction extracts lesion patches from suitable positions within the 3D dataset that serve as input data for the Siamese network. We performed several experiments with different multi-channel patch types, inserting combinations of PET, CT or segmentation data into either a 2D or 3D Siamese architecture.

\section{Methods}\label{sec2}



\subsection{Workflow Overview}\label{ProjectStructure}

Each baseline and follow-up full-body CT and PET scan in DICOM format was processed by the qPSMA software\cite{Gafita2019}, to create a lesion segmentation mask. This algorithm extracts bone lesions by applying an SUV threshold within a segmented bone mask, taking into account a possible misalignment of PET and CT data.

As the patient's position can be shifted in follow-up scans, a rough alignment is required for relocalization during later patch extraction. Therefore an affine registration is applied (section \ref{ImageRegistration}).

In a next step image patches are cropped around every lesion and its surrounding region in the baseline and the follow-up scan. The suitable extraction positions are determined by a case distinction algorithm (section \ref{PatchExtraction}).

The Siamese CNN (section \ref{SiameseCNN}) compares pairs of baseline and follow-up lesion patches and decides, whether the patch pair shows corresponding lesions. The approach narrows the tracking down to a classification problem: true patch pair for corresponding lesions, false patch pair for non-corresponding lesions.

We performed our experiments in two and three dimensions. Therefore, we created a related 2D and 3D Siamese network and respectively cropped all patches in a 2D and 3D version.\\


\begin{figure}[H]
	\centering
	\includegraphics[width=1.0\textwidth]{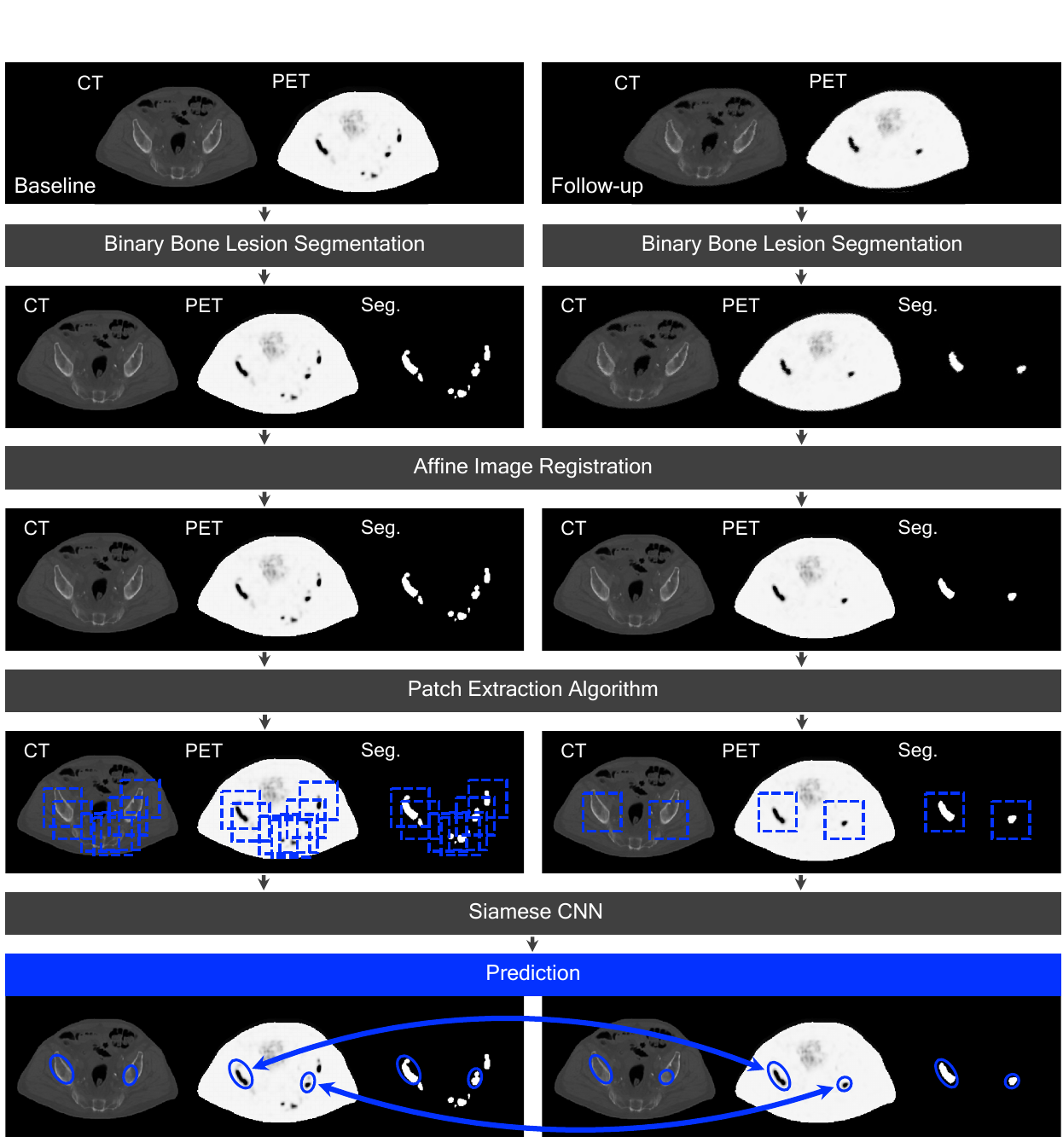}	
	\caption{\textbf{Processing workflow:} Schematic presentation of the proposed pipeline for AI-based PET/CT lesion tracking. Data preparation of the baseline and follow-up scans include a binary bone lesion segmentation (Seg.) and an affine image registration. Patches are extracted around every lesion and then analyzed by a siamese CNN, which assigns the lesions in the baseline and follow-up PET/CT scans.}
	\label{fig:ProjectOverview}
\end{figure}




\subsection{Image Registration}\label{ImageRegistration}

Since the body shape of the patients possibly changes in the course of therapy, an affine registration is performed in the pipeline that also allows for shearing and scaling in addition to the rotation and translation of a rigid transformation. The follow-up CT scan is registered towards the baseline CT scan. As a result we obtain a shear and rotation matrix $ \mathbf{A} = (a_{11}, ...\  a_{33} ) $ and the translation vector $ \mathbf{t} = (t_1, t_2, t_3) $, along with the center of rotation $ \mathbf{c} =  (c_1, c_2, c_3) $, defined as the center of the baseline scan. The transformation is then applied on the follow-up PET scan as well as on its 3D bone segmentation mask.

For the relocalization during patch extraction, positions $ \mathbf{x} $ in the baseline and follow-up original scans can be projected into the respective other scan using the forward transformation $ \mathbf{T}(\mathbf{x}) $ or the reverse transformation $ \mathbf{T^{-1}}(\mathbf{x}) $ by applying the inverse shear and rotation matrix $ \mathbf{A}^{-1} $.

\begin{equation}
	\mathbf{T}: \ \mathbf{x}_{follow-up} = \mathbf{A} \cdot \left( \mathbf{x}_{\ baseline} - \mathbf{c} \right) + \mathbf{t} + \mathbf{c}
\label{eq:TransferBaslinetoFollowUp}
\end{equation}

\begin{equation}
	\mathbf{T^{-1}}: \ \mathbf{x}_{\ baseline} = \mathbf{A}^{-1} \cdot \left( \mathbf{x}_{follow-up} - \mathbf{c} - \mathbf{t}  \right) + \mathbf{c}
\label{eq:TransferFollow-UptoBaseline}
\end{equation}

For image registration SimpleElastix\cite{Marstal2016} was used.

\subsection{Patch Extraction}\label{PatchExtraction}
\begin{figure}[h]
	\centering
	\includegraphics[width=1.0\textwidth]{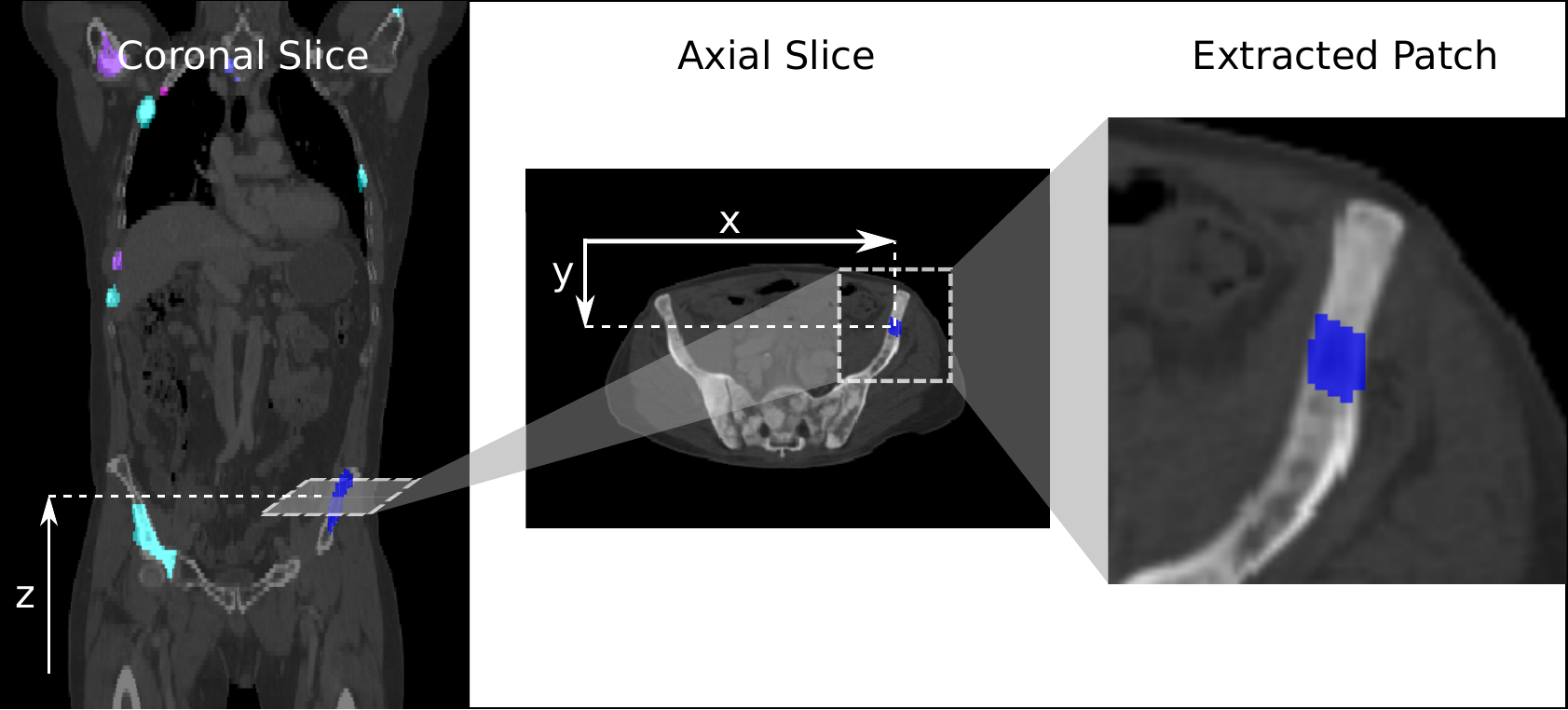}
	\caption{\textbf{2D lesion patch extraction:} An axial 50$\times$50 pixel patch is cropped around the determined extraction point $p(x,y,z)$.}
	\label{fig:PatchExtraction}
\end{figure}

The patches in this study were obtained by cropping axial layers from a whole-body dataset. For a 2D patch around a point of interest $p(x,y,z)$, the z-position selects the axial layer of the dataset, which consists of a CT scan, PET scan, and binary lesion segmentation. The scans are cropped around the x-y-position within a 50$\times$50 pixel frame, as shown in figure \ref{fig:PatchExtraction}. In 3D, several axial slices are included in the patch. For the experiments, 3D patch sizes of 50$\times$50$\times$5 pixels and 50$\times$50$\times$11 pixels were chosen. The extraction position is determined in the same way as for 2D patches, with the only difference being that for 50$\times$50$\times$5 pixel patches, two axial layers above and below the determined z-position are added to the patch, and for 50$\times$50$\times$11 pixel patches, respectively, five axial layers above and below. As described in \ref{Dataset}, the resolution in x- and y-direction is higher than in z-direction and thus a pixel in an x-y-plane displays a smaller area than those in z-direction. 
\\

\begin{figure}[H]
	\centering
	\includegraphics[width=1.0\textwidth]{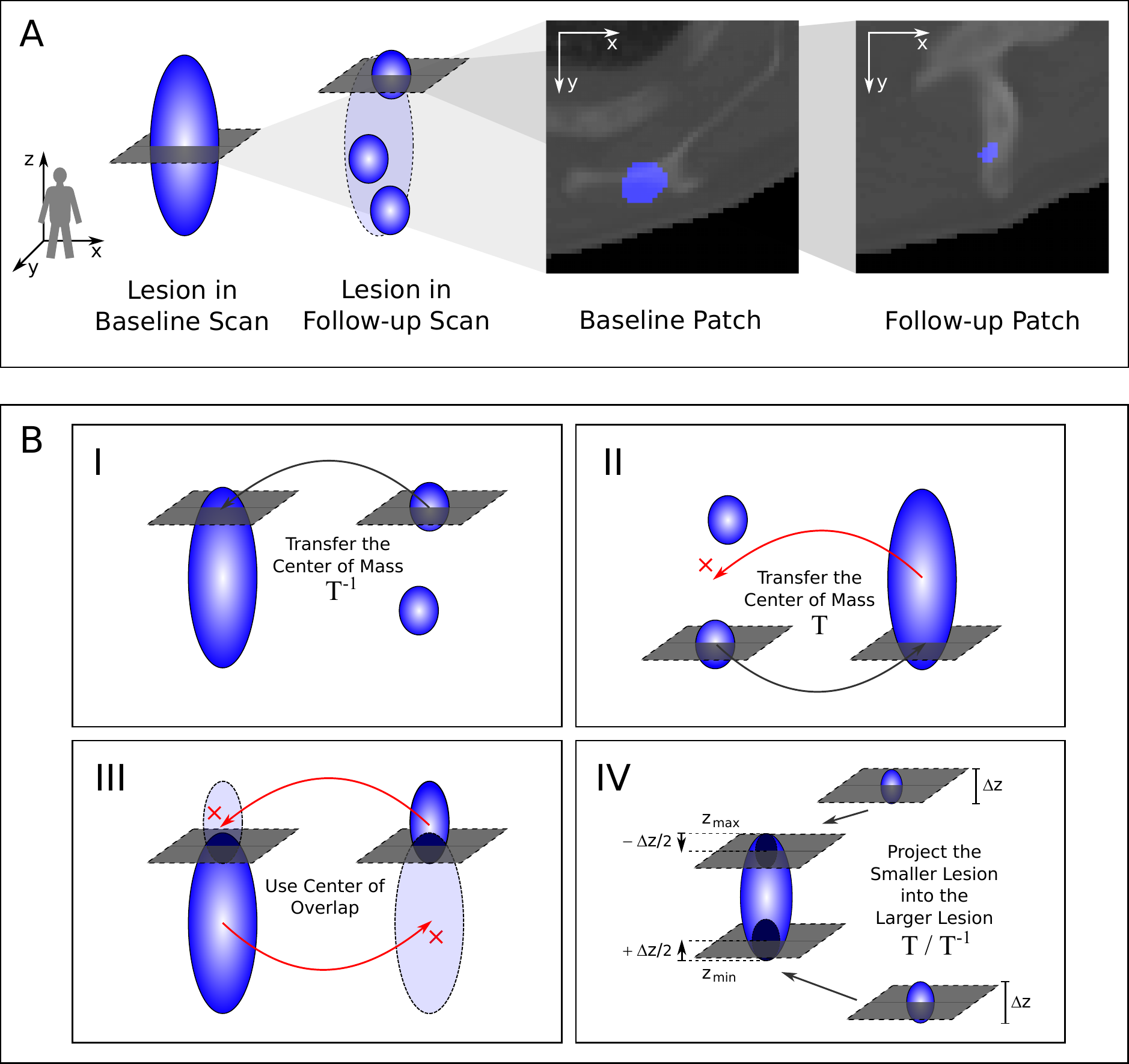}	
	\caption{\textbf{Principle of the patch extraction algorithm:}\\
		\textbf{(A) Unsuitable patch pair after extraction at center of mass:} A large baseline lesion shrank and divided into three smaller lesions in the follow up-scan. The example shows a baseline and follow-up lesion patch of the left scapula extracted at the lesions' center of mass (CoM). Patch extraction at each lesion's CoM can lead to patch pairs showing different anatomical environments, even though they show corresponding lesions. For this reason, a detailed patch extraction algorithm determines the suitable patch extraction point $p(x,y,z)$.\\
		\textbf{(B) Patch extraction cases:} In the algorithm a case distinction hierarchically applies the four cases A-D to find suitable points $p(x,y,z)$ within the baseline and follow-up lesion for the patch pair extraction. If one case does not lead to a result, the algorithm passes on to the next one. For each case, the illustration shows the baseline lesion(s) on the left side and the follow-up lesion(s) on the right side. The gray cross-sectional areas indicate the final extracted axial patches. $ \mathbf{T} $ and $\mathbf{T^{-1}}$ indicate point transfers to the respective other scan (eq. \ref{eq:TransferBaslinetoFollowUp}, eq. \ref{eq:TransferFollow-UptoBaseline}). }
	
	\label{fig:UnsuitablePatchPair+Cases}
\end{figure}

When creating a two-dimensional or a small three-dimensional patch out of a three-dimensional lesion, the position of the extraction is of high importance. Extracting the patch only at the Center of Mass (CoM) of each lesion and then creating patch pairs can lead to unsuitable true patch pairs that do not show the same anatomical region, even though they represent corresponding lesions. This is the case when a baseline lesion divides into several small follow-up lesions, as shown in figure \ref{fig:UnsuitablePatchPair+Cases}A. When comparing a relatively small to a large lesion, the two centers of mass are on different axial layers, even though they are corresponding lesions. Therefore, the ROI of the larger lesion, within the comparison with the smaller one, is not its CoM. Instead, its ROI is represented by the anatomical environment of the smaller lesion. It is, thus, necessary to transfer the z-coordinate of the second CoM into the large lesion and extract a patch around the larger lesion at the adjusted axial layer. Hence, the lesion patches cannot be extracted independently. It is always necessary to extract a patch pair together, as the patch position for a lesion depends both on its own structure and on that of the compared lesion.
\\


An algorithm uses hierarchical cases applied sequentially to determine the patch extraction point (figure \ref{fig:UnsuitablePatchPair+Cases} B). The 3D CoM of a lesion is used as a starting point and centered within the 2D axial slice of the lesion.

\textit{Case I} projects the corrected CoM of the follow-up lesion into the baseline lesion using $\mathbf{T^{-1}}(\mathbf{x})$, mostly used if the baseline lesion is larger and has either shrunk in volume or has divided into several smaller follow-up lesions.

\textit{Case II} is implemented if Case I is not successful, projecting the corrected CoM of the baseline lesion into the follow-up lesion ($\mathbf{T}(\mathbf{x})$), mostly applied with the baseline lesion being the smaller one.


\textit{Case III} is utilized if due to shifts between the baseline and follow-up lesions the transferred CoMs cannot be identified within the lesions, mostly due to minor registration errors. To address this challenge, the algorithm locates the intersection of both lesions and designates the center of the overlap as the patch point for each lesion. The overlap is then transformed using $\mathbf{T}(\mathbf{x})$ to determine its coordinates in the follow-up scan.

Even though Cases A to C mostly extract true lesion pairs, sometimes, due to registration errors, they may yield false patch pairs. Therefore, the use of the Siamese CNN approach remains relevant.

\textit{Case IV} is designed for lesions that do not overlap, which can happen when corresponding lesions are not accurately registered. This method is also utilized for most false patch pairs. For the smaller lesion in the pair, the patch point is set at the axial layer of its CoM. For the larger lesion, the shape of the smaller lesion is projected onto the nearest end of the larger lesion in the z-direction. The point is then taken at the layer where the CoM of the projected smaller lesion would be positioned. Mathematically, the z-coordinate of the extraction point of the larger lesion $z_{extr, l}$ can be expressed as follows:

\begin{equation}
	z_{extr, l} = \begin{cases}
	z_{max, l} - \Delta z_{s}/2 & \ \  \text{if} \ z_{CoM, s} > z_{CoM, l}\\
	z_{min, l} + \Delta z_{s}/2 & \ \  \text{if} \ z_{CoM, s} < z_{CoM, l}   \text{  .}
	\end{cases}
\end{equation}

Here, $\Delta z_s$ represents the height in the z-direction of the smaller lesion, and $z_{max, l}$ or $z_{min, l}$ indicates the z-coordinate of the highest or lowest point of the larger lesion. $z_{CoM, s}$ and $z_{CoM, l}$ denote the z-coordinates of the CoMs of the two lesions.\\

For the test set, a lesion from the baseline scan is compared to all follow-up lesions in a specific ROI around the margins of the projected baseline lesion ($ \mathbf{T}(\mathbf{x}) $, eq. \ref{eq:TransferBaslinetoFollowUp}). We defined it with $\Delta x = 10$, $\Delta y = 10$ and $\Delta z = 5$. In z-direction the interval is smaller as its pixel size is usually larger than in x- and y- direction. This ROI was chosen as it contains 99\% of the corresponding lesions, when tested on the train and validation set.

\subsection{Siamese CNN}\label{SiameseCNN}

\begin{figure}[h]
	\centering
	\includegraphics[width=0.7\textwidth]{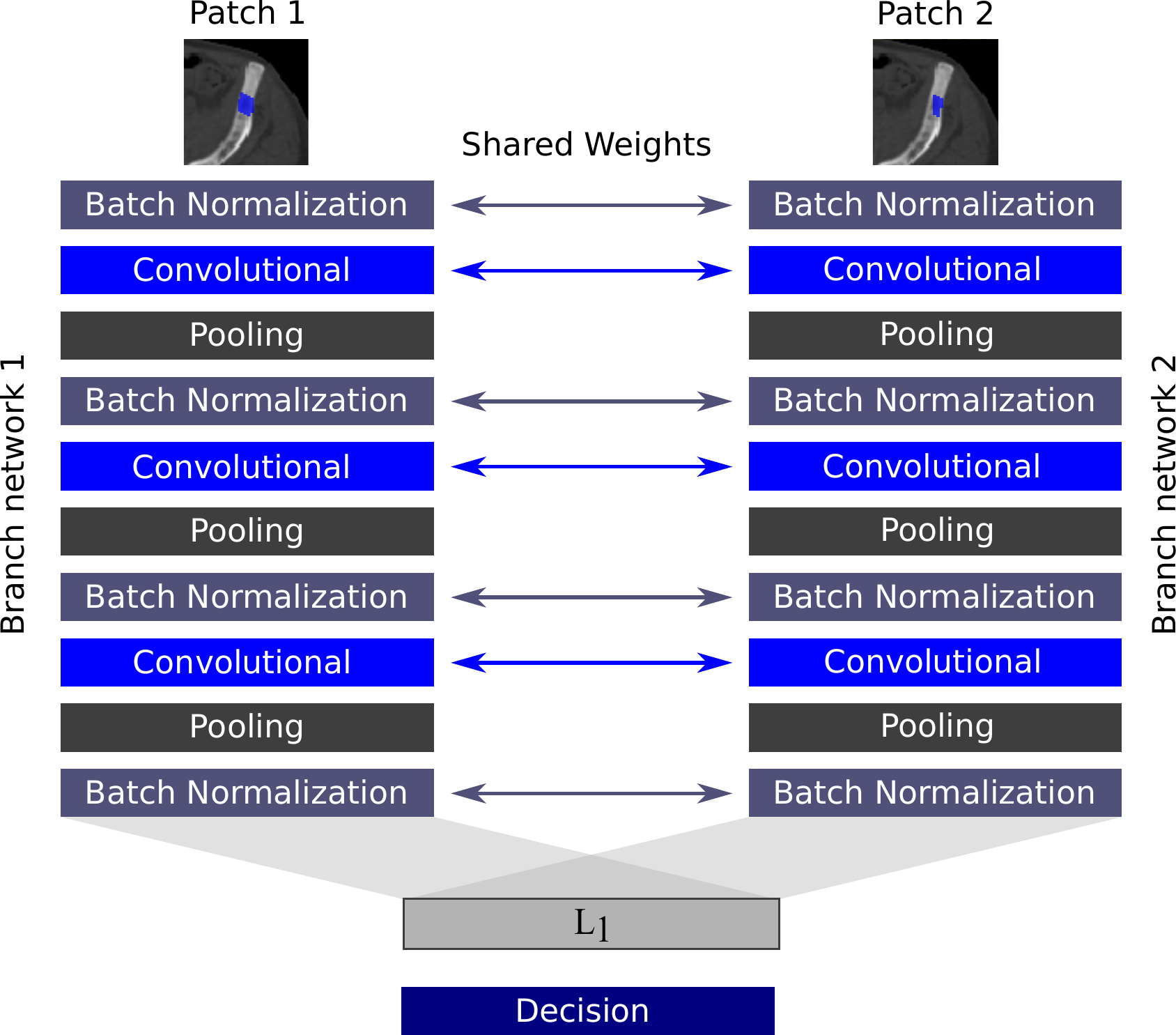}	
	\caption{\textbf{Structure of the siamese network:} Patches are processed by two parallel CNN branches with shared weights, whose output is merged by a $ \mathbf{L_1} $-layer. Architecture details of the branches are shown in table \ref{tab:SiameseNetwork}.}
	\label{fig:SiameseNetwork}
\end{figure}

A Siamese network consisting of two parallel CNN branches was chosen to process the input of patch pairs (fig. \ref{fig:SiameseNetwork}). It uses shared weights for the two branches and was inspired by Koch et al. \cite{Koch2015SiameseNN}, who used a similar network for one-shot image recognition. . Each CNN branch has three convolutional layers with \textit{ReLU} activation and three pooling layers. Batch normalization is applied before every convolutional ayer. The output of a branch again undergoes batch normalization, is flattened to a vector $ \mathbf{h}(\mathbf{X}) $ and is merged in a custom-defined $ \mathbf{L_1} $-layer, which takes the element-wise absolute difference of the feature vectors $ \mathbf{h}(\mathbf{X_1}) $ and $ \mathbf{h}(\mathbf{X_2}) $:

\begin{equation}
\mathbf{L_1}(\mathbf{X_1}, \mathbf{X_2})_{i}= || \mathbf{h}(\mathbf{X_1})_i - \mathbf{h}(\mathbf{X_2})_i ||_1   \text{  .}
\label{eq:L1Layer}
\end{equation}

The output $ \mathbf{L_1}(\mathbf{X_1}, \mathbf{X_2}) $ is fed to a final fully-connected decision layer (dense layer) and a \textit{softmax function} distributing the final network output to the two classes \textit{true patch pair} and \textit{wrong patch pair}.\\

To process the different patch dimensions (2D, 3D), the Siamese network was created in 2D and 3D. Also, for the differently sized 3D patches with z=5 and z=11, the kernel size of the CNN had to be adapted as a larger patch can be further scaled down by larger kernels. Details of the branch architecture are displayed in table \ref{tab:SiameseNetwork}. 



\begin{table}[h]
	\caption{Architecture Details of the used Siamese Branches }\label{tab:SiameseNetwork}
	\begin{tabular*}{\textwidth}{@{\extracolsep\fill}lcccccc}
		\toprule%
		\textbf{Layer} & \textbf{Size} & \textbf{Size} & \textbf{Kernels} & \textbf{Stride} & \textbf{Output} & \textbf{Output}\\
		 & \textbf{2D} & \textbf{3D z5$|$z11} & & \textbf{2D (3D)} & \textbf{2D} & \textbf{3D z5$|$z11} \\
		\midrule
		Input& & & & & 50$\times$50$\times$1\text{*} & 50$\times$50$\times$5$|$11$\times$1\text{*} \\
		Convolutional & 5$\times$5$\times$1\textdagger & 5$\times$5$\times$2$|$4$\times$1\textdagger & 20 & 1$\times$1($\times$1) & 46$\times$46$\times$20 & 46$\times$46$\times$4$|$8$\times$20\\
		Pooling & 3$\times$3 & 3$\times$3$\times$1$|$2 & -- &  2$\times$2($\times$1) & 22$\times$22$\times$20 & 22$\times$22$\times$4$|$7$\times$20 \\
		Convolutional & 3$\times$3$\times$20 & 3$\times$3$\times$2$|$3$\times$20 & 40 & 1$\times$1($\times$1) & 22$\times$22$\times$40 & 22$\times$22$\times$3$|$5$\times$40 \\
		Pooling & 2$\times$2 & 2$\times$2$\times$1$|$2 & -- &  2$\times$2($\times$1) & 11$\times$11$\times$40 & 11$\times$11$\times$3$|$5$\times$40 \\
		Convolutional & 3$\times$3$\times$40 & 3$\times$3$\times$2$|$3$\times$40 & 50 &  1$\times$1($\times$1) & 9$\times$9$\times$50 & 9$\times$9$\times$2$|$3$\times$50 \\
		Pooling & 2$\times$2 & 2$\times$2$\times$1$|$2 & -- &  2$\times$2($\times$1) & 4$\times$4$\times$50 & 4$\times$4$\times$2$|$2$\times$50 \\
		\botrule
	\end{tabular*}
	\footnotetext{The table shows the structure of the layers for a 2D Siamese branch and for a 3D Siamese branch, which exists in two different versions for patches with z=5 and patches with z=11. Annotation example for the kernel size of the first 3D convolutional layer: 5$\times$5$\times$2$\times$1 for patches with z=5 and 5$\times$5$\times$4$\times$1 for patches with z=11.}
	\footnotetext{\text{*}1-channel patches as input. For 2-channel patches, the input shape is 50$\times$50$\times$2 (2D) and 50$\times$50$\times$5$|$11$\times$2 (3D), respectively.}
	\footnotetext{\textdagger First layer structure for 1-channel patches. For 2-channel patches, only the kernel size of first layer is extended to 5$\times$5$\times$2 (2D)and 5$\times$5$\times$4$|$2$\times$2 (3D), respectively.}
\end{table}

Applying the \textit{Adam} optimizer\cite{Adam}, which already effectively adapts the learning rate due to momentum optimization, for our network case, we achieved best performance with an initial learning rate of $ 1 \times 10^{-4} $. Nevertheless, an additional combination of learning rate scheduling together with momentum optimization has shown good results.\cite{Senior2013} In several siamese networks, different learning rate decay have been applied.\cite{Koch2015SiameseNN}\cite{Zagoruyko2015} For this reason, we chose \textit{performance scheduling} for the learning rate with a patience of $ 5 $ epochs, a scheduling factor of $ 0.2 $ and a minimum learning rate of $ 1 \times 10^{-5} $.

As the lesion tracking task requires the comparison of different scans with possibly changed morphologies during therapy, a high grade of generalization is necessary. In order to achieve this and prevent overfitting, regularization is applied in various ways. In addition to the batch normalization, $ \ell^2 $\textit{-parameter regularization} is used with a penalty factor of $ 5 \times 10^{-4} $ as recommended for CNNs.\cite{AlexNet2017} Following further suggestions for CNNs, a dropout rate of 40\% is applied to the top three layers.\cite{Szepesi2022}

\section{ Experiments and Results}

\subsection{Dataset }
\label{Dataset}

In this project, the processed dataset consists of 36 patients with metastatic castration resistant prostate cancer with a high number of bone lesions requiring follow-up tracking. The patients were treated with [$^{177}$Lu]Lu-PSMA-I\&T (imaging and therapy)\cite{Karimzadeh2023} and a total number of 188 cycles (median 5 cycles, range 1-31). Eligibility criteria were previous treatment with abiraterone or enzalutamide, previous taxane-based chemotherapy or chemoineligibility, and positive PSMA-ligand uptake at PET scan. The [$^{177}$Lu]Lu-PSMA-I\&T was given 6-8 weekly with an activity of 7.4 GBq in up to six cycles in a row with a potential continuation after a therapy break. Within the selected patient group, prostate-specific antigen decline of $>50\%$ was achieved in 15 patients (43 \%), median PSA-progression-free survival was 5.3 months, and median overall survival (OS) was 14.2 months. All patients gave written consent for evaluation of their data.

For each patient a full-body baseline and a follow-up [$^{68}$Ga]Ga-PSMA-11\cite{Eder2012} or [$^{18}$F]F-rhPSMA-7\cite{Wurzer2020}\cite{Wurzer2020a} PET/CT scan after two cycles of [$^{177}$Lu]Lu-PSMA-I\&T\cite{Karimzadeh2023} therapy are analyzed, adding up to 2111 baseline lesions and 2658 separately segmented foci in the follow-up scans. The baseline and follow-up lesions show 1490 true lesion assignments.

The PET/CT tracers were administered in compliance with the German Medicinal Products Act, AMG x13(2b), and in accordance with the responsible regulatory body (Government of Oberbayern). They were synthesized and all scans performed at our institution as decribed in Eiber et al.\cite{Eiber2015} and Kroenke et al. \cite{Kroenke2022}. For the image acquisition the scanners Siemens Biograph mCT and Siemens Biograph Vision were used. After image reconstruction, the transaxial pixel size was 4.07 mm for PET and 1.52 mm for CT, each with a 5 mm or 3mm slice thickness.

Ground truth lesion pairs were manually assigned by an experienced nuclear medicine physician using an in-house software. The dataset was splitted into 70\% training set, 15\% validation set and 15\% test set. The distribution was applied in a way that ensured this ratio for the patients as well as for the lesions. In addition, data augmentation in form of patch rotation was implemented to further increase the size of the training dataset.

\subsection{Experimental Setup}

Lesions may evolve along different patterns, which have to be recognized by the pipeline. A lesion can be remaining in both scans, split into several lesions or fuse with neighboring lesions. Also, a baseline lesion can resolve completely or a new lesion can develop in the follow-up.

The Siamese CNN is trained with balanced ground truth data containing equal numbers of true patch pairs and false patch pairs. For the application of the network on a test set, all possibly relevant lesion combinations are fed to the CNN in the shape of patch pairs. The predictions are used to assign the lesions. If all patch pairs of a baseline lesion are classified as false patch pairs, this lesion can be declared as resolved and no follow-up lesions are assigned. The same situation for a lesion from the follow-up scan suggests that it is a new lesion.

As described above (sec. \ref{PatchExtraction} and \ref{SiameseCNN}), experiments have been performed with a 2D network and two 3D network variations. Respectively, all extracted patches exist in a 2D and two 3D versions.

Since whole-body imaging data is available from CT, PET and binary lesion segmentation, all of it can serve as an input for the patch extraction. Therefore, we processed four different patch types in the network, single-channel CT patches and several two-channel patches that combined CT \& PET, CT \& Binary Lesion Segmentation and CT \& Segmented CT data. As \textit{Segmented CT} we define the element-wise product of the Binary Lesion Segmentation patch with the CT patch, showing the CT information only within the segmented lesion.

All runs have been carried out several times with different seeds to show statistical stability.

\subsection{Reference Model}\label{ReferenceModel}
To validate the learning ability of the used network, the performance of the trained CNN is compared to that of a non-trainable model, which only compares the intensities of two 2D CT patches by subtracting them and taking the pixel-wise absolute value of the difference. This is equal to directly applying the $ \mathbf{L_1} $-layer (eq. \ref{eq:L1Layer}) without any CNN branches. The reference model is incapable of learning to compare anatomical structures. The decision output $ y $ of the intensity model with the input patches $ \mathbf{X_1} $ and $ \mathbf{X_2} $ of shape $ (N, N) $ can be described as:

\begin{equation}
y = 1 - \frac{1}{N^2} \sum_{i,j = 1}^{N} \mathbf{L_1}_{i,j}(\mathbf{X_1}, \mathbf{X_2})
\end{equation}

\subsection{Siamese Network Performance}

Every training was carried out for 400 epochs. We chose the lowest validation loss as  criteria to select the epoch with the best weights. During testing, as default, a threshold of 0.5 is applied for the classification problem with the two classes true patch pair and false patch pair. For each combination of 2D or 3D Siamese CNN and the different patch types, table \ref{tab:NetworkPerformance} shows the performance result on the validation set during training and on the test set. 

\begin{table}[h]
	\caption{Performance of the Siamese network}\label{tab:NetworkPerformance}
	\begin{tabular*}{\textwidth}{@{\extracolsep\fill}lcccccc}
		\toprule%
		\textbf{Dim.} & \textbf{Patch Size} & \textbf{Patch Type} & \textbf{Validation Acc} & \textbf{Validation Loss} & \textbf{Test Accuracy}  \\
		\midrule
		2D & 50$\times$50 & CT & $ 0.790 \pm 0.015 $ & $ 0.508 \pm 0.030 $ &  $ 0.830 \pm 0.003 $  \\ 
		2D & 50$\times$50 & CT/PET & $ 0.800\pm 0.008 $ & $ 0.521 \pm 0.016 $ &  $ 0.806 \pm 0.012 $  \\ 
		2D & 50$\times$50 & CT/Seg & $ 0.781 \pm 0.010 $ & $ 0.530 \pm 0.011 $ &  $ 0.795 \pm 0.001 $  \\ 
		2D & 50$\times$50 & CT/Seg CT & $ 0.795 \pm 0.011 $ & $ 0.515 \pm 0.013 $ &  $ 0.788 \pm 0.011 $  \\ 
		
		3D & 50$\times$50$\times$5 & CT & $ 0.804 \pm 0.008 $ & $ 0.486 \pm 0.008 $ &  $ 0.820 \pm 0.008 $  \\ 
		3D & 50$\times$50$\times$5 & CT/PET & $ 0.802 \pm 0.002 $ & $ 0.499 \pm 0.007 $ &  $ 0.786 \pm 0.019 $  \\ 
		3D & 50$\times$50$\times$5 & CT/Seg & $ 0.819 \pm 0.005 $ & $ 0.472 \pm 0.010 $ &  $ 0.790 \pm 0.022 $  \\ 
		3D & 50$\times$50$\times$5 & CT/Seg CT & $ 0.821 \pm 0.011 $ & $ 0.475 \pm 0.023 $ &  $ 0.803 \pm 0.012 $  \\ 
		
		3D & 50$\times$50$\times$11 & CT & $ 0.829 \pm 0.013 $ & $ 0.441 \pm 0.016 $ &  $ 0.786 \pm 0.007 $  \\ 
		3D & 50$\times$50$\times$11 & CT/PET & $ 0.817 \pm 0.004 $ & $ 0.485 \pm 0.006 $ &  $ 0.761 \pm 0.017 $  \\ 
		3D & 50$\times$50$\times$11 & CT/Seg & $ 0.830 \pm 0.004 $ & $ 0.463 \pm 0.007 $ &  $ 0.791 \pm 0.017 $  \\ 
		3D & 50$\times$50$\times$11 & CT/Seg CT & $ 0.836 \pm 0.007 $ & $ 0.465 \pm 0.016 $ &  $ 0.790 \pm 0.018 $  \\ 
		\botrule
	\end{tabular*}
	\footnotetext{Seg: Binary Lesion Segmentation; Seg CT: Segmented CT }
\end{table}

On the test set, the 2D Siamese CNN outperforms for most of the patch types. Within the same network type, training with the single-channel CT patches reaches the best test accuracy, especially for the 2D network (83 \%) and the 3D Siamese CNN with 50$\times$50$\times$5 patches (82 \%). Considering the standard deviation, the CT patches also work amongst the best in the 3D network with 50$\times$50$\times$11 patches.

For the 3D networks training, the two-channel CT/Segmented CT patches rank second and yield good test accuracy values with e.g. 80\% for 50$\times$50$\times$5 patches. Furthermore, the direct comparison between the two 3D Siamese networks shows better test results for the 50$\times$50$\times$5 architecture.

\begin{figure}[h]
	
	\begin{subfigure}[c]{0.5\textwidth}
		\centering
		\includegraphics[width=1.0\textwidth]{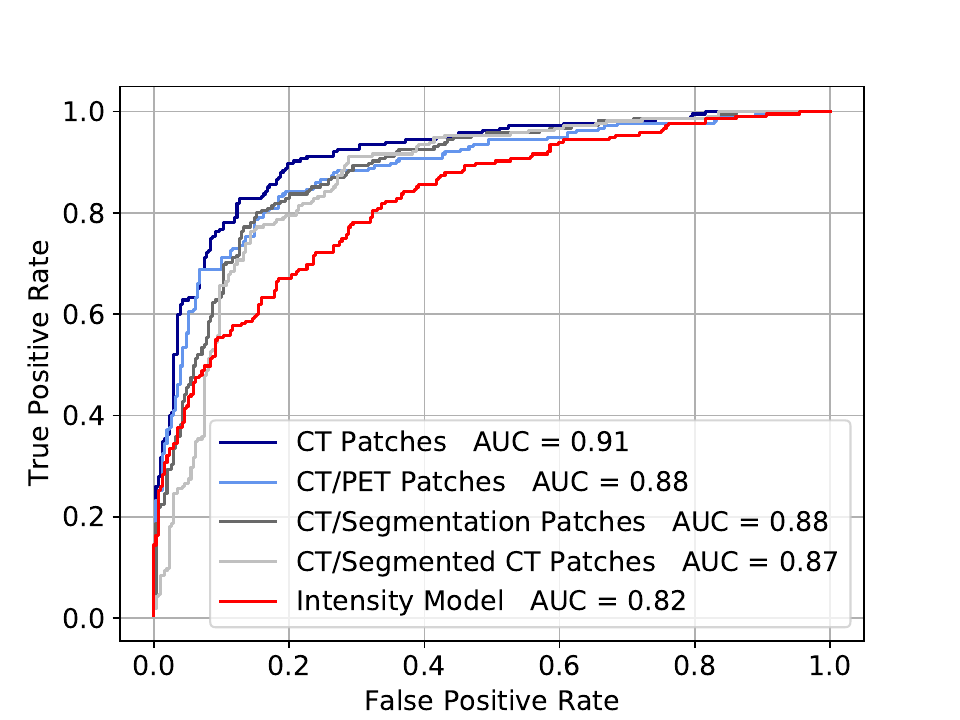}
		\subcaption{2D Siamese CNN with 50$\times$50 Patches}
	\end{subfigure}
	\begin{subfigure}[c]{0.5\textwidth}
		\centering
		\includegraphics[width=1.0\textwidth]{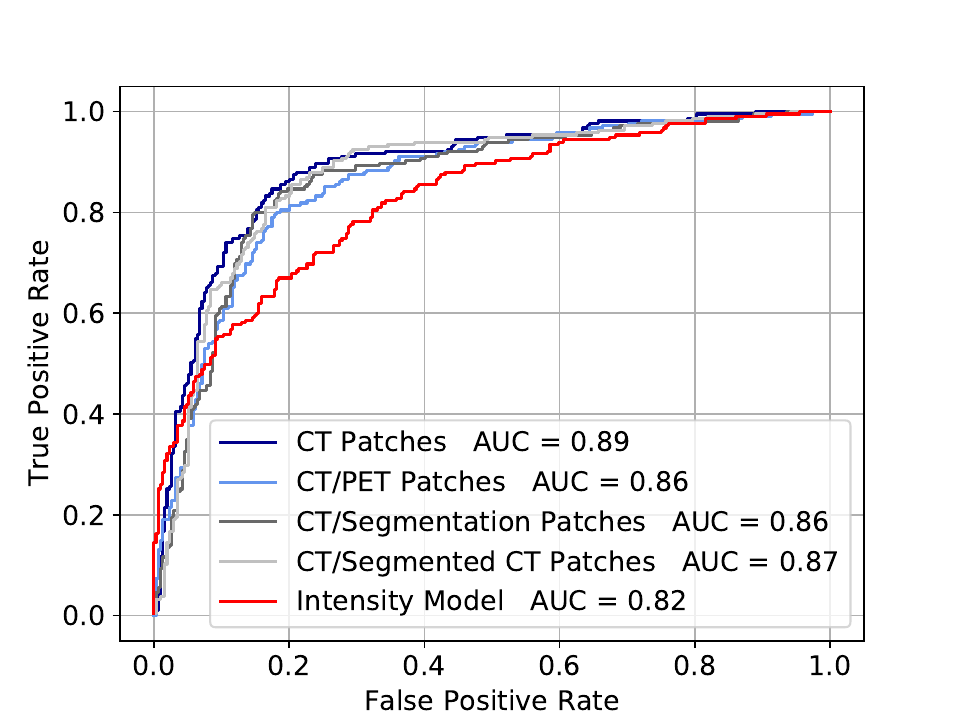}
		\subcaption{3D Siamese CNN with 50$\times$50$\times$5 Patches}
	\end{subfigure}	
	\begin{subfigure}[c]{0.5\textwidth}
		\centering
		\includegraphics[width=1.0\textwidth]{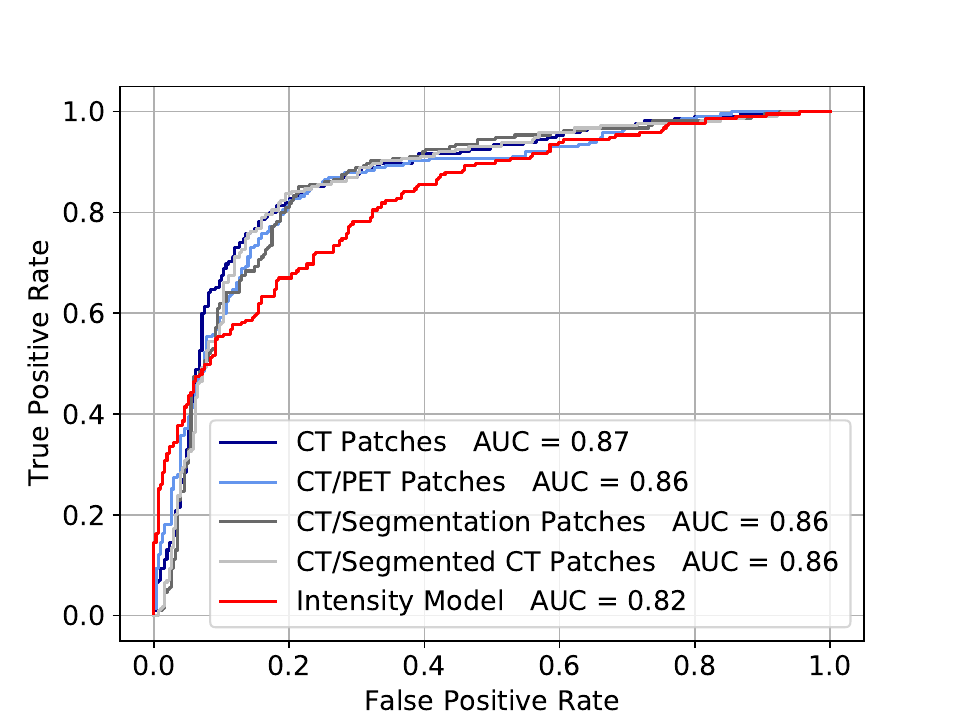}
		\subcaption{3D Siamese CNN with 50$\times$50$\times$11 Patches}
	\end{subfigure}
	
	\caption{\textbf{ROC curves of the siamese networks trained with different patch types:} The AUC values for each curve are given in the legend. The best network configuration is reached with a 2D Siamese network trained with single-channel CT patches (AUC=0.91). The non-trainable reference intensity model (sec. \ref{ReferenceModel}) reaches an AUC=0.82, which is outperformed by all networks and patch types.}
	\label{fig:ROC}
\end{figure}

Since the accuracy depends on the applied threshold of the network, we analyzed the results with ROC curves and the respective AUC. The ROC curves of the differently trained networks are shown in figure \ref{fig:ROC} and confirm the described results. As a reference model, all graphs show the ROC of the non-trainable intensities model (sec. \ref{ReferenceModel}). All trained networks outperform the intensity model proving a successful training process. Clearly, the 2D Siamese network trained with single-channel CT patches reaches the best result with an AUC=0.91.


\begin{figure}[h]
	\centering
	\includegraphics[width=0.6\textwidth]{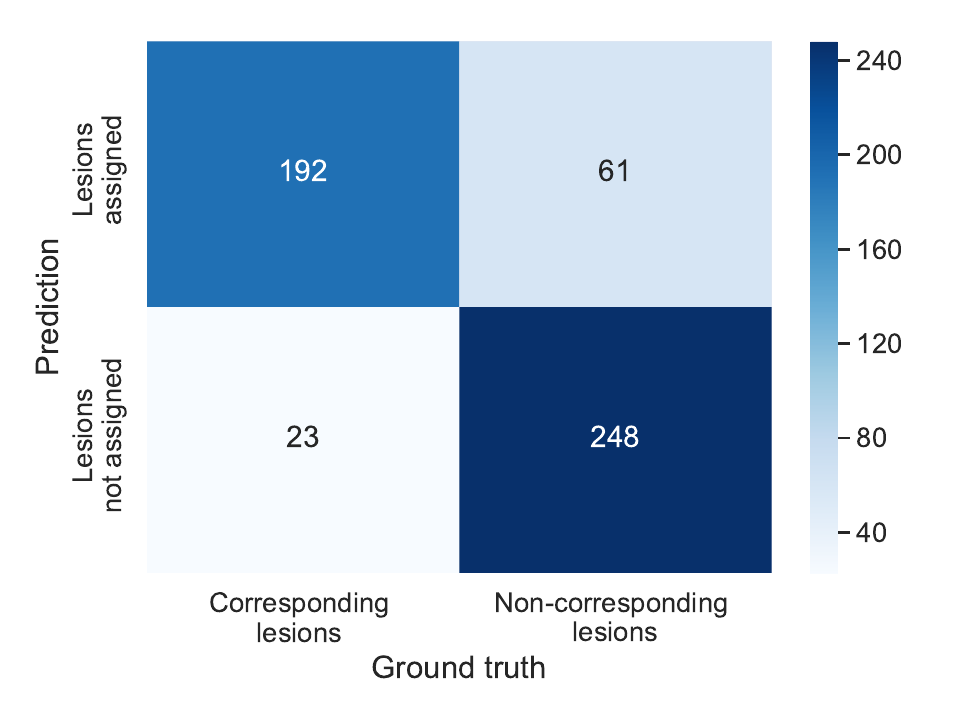}	
	\caption{\textbf{Confusion matrix for lesion tracking:} The figure illustrates the number of true positives (upper left), false positives(upper right), false negatives (lower left) and true negatives (lower right) among a cumulative total of 524 decisions by the 2D siamese network trained with 50$\times$50 CT patches. Threshold optimization was performed by $ G_{\text{mean}} $ maximization.}
	\label{fig:Confusion Matrix}
\end{figure}

To obtain the best model indicated by the upper left of the ROC, we performed threshold optimization by maximizing $ G_{\text{mean}} $. Table \ref{tab:StatisticalParameters} displays the statistical parameters before and after threshold optimization. The final model reaches an 83\% accuracy, a 88\% sensitivity, a 80\% specificity and a 76\% precision. This results in a $ G_{\text{mean}} $ of 0.84 and a $ F_1 $-score of 0.82. The resulting confusion matrix with a total number of 524 decision cases in the test set is shown in figure \ref{fig:Confusion Matrix}.\\

\begin{table}[h]
	\caption{Statistical parameters for the performance of the 2D Siamese network trained with 50$\times$50 CT patches}\label{tab:Statistics}
	\begin{tabular}{@{}lllll@{}}
		\toprule
		\textbf{Threshold} & \textbf{Accuracy}  & \textbf{Sensitivity} & \textbf{Specificity} & \textbf{Precision} \\
		\midrule
		Default 0.5    & $ 0.830 \pm 0.003 $   & $ 0.731 \pm 0.039 $  & $ 0.910 \pm 0.032 $ & $ 0.850 \pm 0.037 $ \\
		$ G_{\text{mean}} $ optimized   & $ 0.833 \pm 0.006 $   & $ 0.881 \pm 0.021 $  & $ 0.800 \pm 0.016 $ & $ 0.755 \pm 0.012 $ \\
		\botrule
	\end{tabular}
	\label{tab:StatisticalParameters}
\end{table}

\begin{figure}[H]
	\centering
	\includegraphics[width=0.6\textwidth]{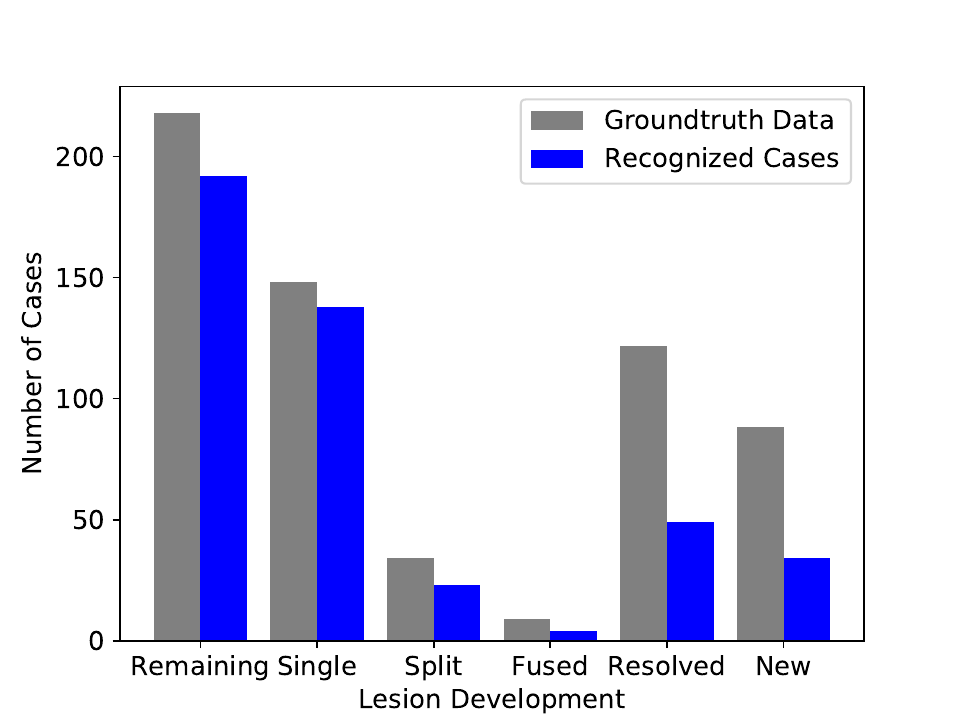}	
	\caption{\textbf{Pipeline performance on different lesion development cases:} Remaining lesion cases, with the subcategories of lesions remaining as single lesions, split lesion cases and fused lesion cases, and finally the cases of resolved (disappeared) lesions and newly developed lesions.}
	\label{fig:PerformanceLesionDevelopment}
\end{figure}

As a last step of the pipeline, the results of the network decisions are used to determine if baseline lesions remain, resolve or newly appear in the follow-up. The pipeline performance on the detection of the different lesion development cases are shown in figure \ref{fig:PerformanceLesionDevelopment}. Remaining lesions are tracked successfully in 89.3\% of the cases. Amongst them, baseline lesions persisting as single lesions in the follow-up are recognized with a 93.2\% success rate, whereas splitting cases are recognized in 67.6\% and fused lesions in 44.4\% of the cases. Resolved baseline lesions are correctly identified in 40.2\% and new follow-up lesions in 38.6\% of the instances.

\begin{figure}[H]
	\centering
	\includegraphics[width=1.0\textwidth]{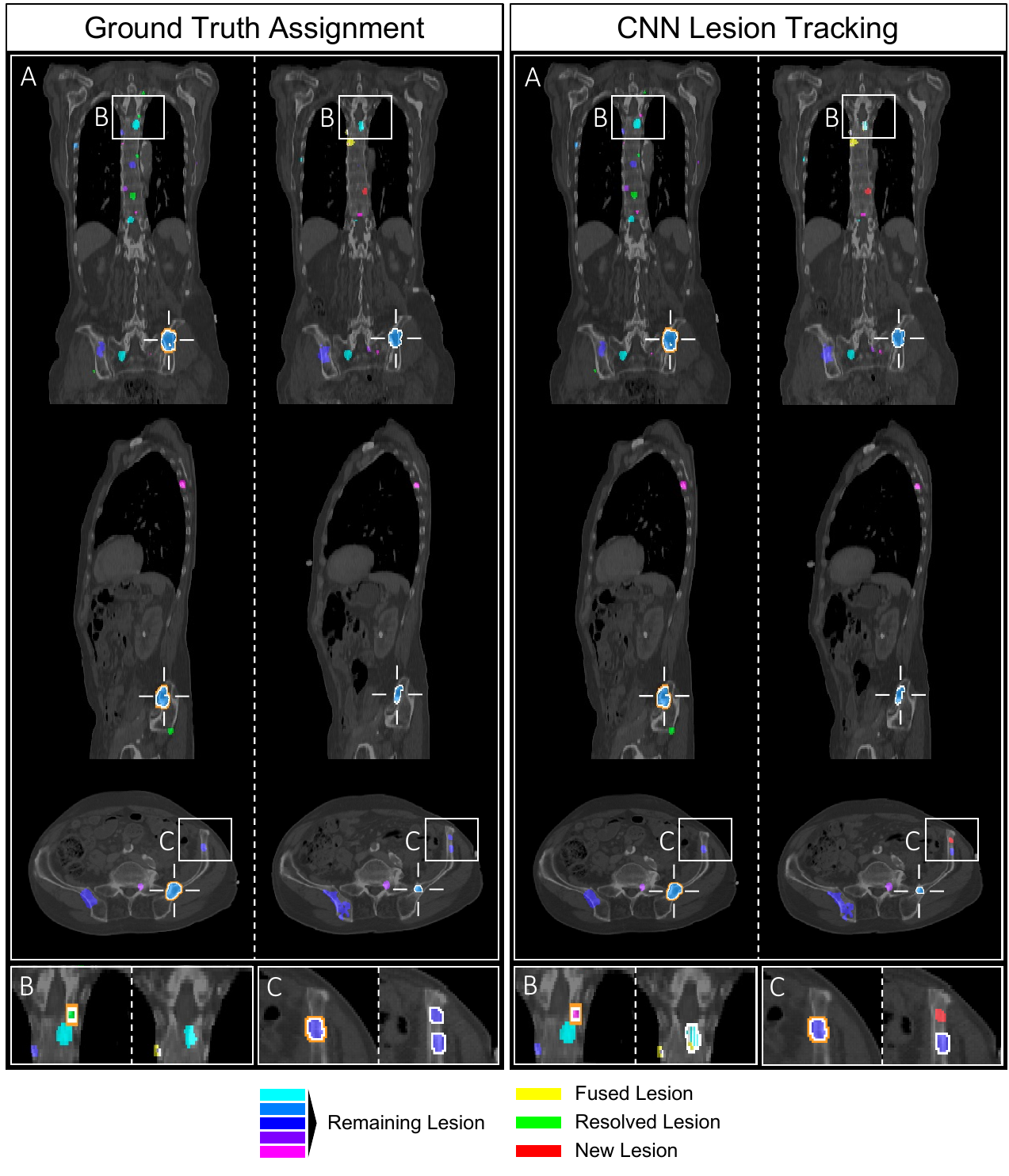}	
	\caption{\textbf{Application of CNN lesion tracking on PSMA-PET/CT scans of a prostate cancer patient:} Comparison of the nuclear medicine specialist's ground truth lesion assignment (left column) to the result of CNN lesion tracking (right column), with each the baseline scan on the left side of the dotted line and the follow-up on the right side. Baseline and follow-up lesions with the same color have been assigned as corresponding lesions, fused lesions are marked yellow, resolved lesions green and new lesions red. (A) 3D overview of the lesion distribution and the tracking result: Exemplary, the coronal, sagittal and axial view display planes of the lesion indicated with the white cross. All shown lesions except one (B) have been correctly tracked by the CNN. (B) Resolved lesion closely located to a remaining lesion: The CNN has wrongly classified the resolved lesion as a fusion with the remaining lesion due to its close anatomic spatial relation. Often, these cases even show a high inter-observer variability amongst nuclear medicine specialists. (C) CNN outperforms nuclear medicine specialist: The ground truth set shows a wrongly annotated split lesion. However, the CNN suggested the second follow-up lesion to be a new lesion. After reviewing the PET dataset, the CNN's decision has been proven correct.}
	\label{fig:ExampleCNNLesionTracking}
\end{figure}

An application example of Siamese CNN lesion tracking in two consecutive scans of a patient is shown in figure \ref{fig:ExampleCNNLesionTracking}, where the CNN in one lesion tracking case even outperforms the nuclear medicine specialist (fig. \ref{fig:ExampleCNNLesionTracking} C).\\

\newpage
\thispagestyle{empty}

\section{Discussion}\label{Discussion}

In our project, for the first time, an AI-based pipeline proved its ability to successfully and automatically track bone lesions in PSMA-PET/CT scans by using the two imaging modalities. The high sensitivity and specificity of the PET enables clear lesion segmentation and positioning, whereas the high anatomic contrast of the CT facilitates accurate position tracking. In contrast to mere CT imaging, where lesions have to be located in the follow-up without knowing their shape in advance, in multimodal PET/CT, the follow-up lesion is already presegmented by the PET and only has to be assigned with the help of its CT position. This is a main advantage for the analysis of the high number of neighboring lesions in the PSMA-PET/CT scans. Our technique models the way of a physician relating baseline and follow-up.

This assumption has been proved in our experiments. The analysis revealed that PET data and its binary lesion segmentation are crucial for localization during patch extraction. However, the PET and segmentation information within patches are not consistent enough for lesion tracking by comparing these patches. This is supported by the fact that CT patches in our pipeline outperformed all other patch types with two channels, including the PET, binary lesion segmentation data, or the segmented CT, which also depends on the PET lesion segmentation. While the anatomical environment of the lesions represented by the CT data is roughly maintained during therapy, the shape of the PET signal at its segmentation can change strongly. This data in a second channel does not provide added information for the Siamese CNN but might even mislead it.

Even with single-channel CT patches for the Siamese network, the overall procedure is a PET/CT lesion tracking approach. Our project differs from other mere CT lesion tracking approaches as the lesions are selected and cropped to patches on the basis of the PET scan instead of the CT scan. In our approach, the position of the CT patches contains the information given by the PET data.\\

Given the lesion tracking task as a three-dimensional problem, the result of our experiments with the best performance achieved by a 2D Siamese CNN instead of a 3D network seems surprising. However, the 3D positional relation of the lesions is already processed by the case distinction algorithm during patch extraction (sec. \ref{PatchExtraction}) passing the information on in form of extracted patches and their lesion IDs. The 3D patches contain more additional information about the environment, which is partly overlapping with the information handled by the case distinction algorithm. Even though the larger patches processed by 3D networks with far more parameters partly reaches similar results as the 2D network, the larger input does not represent an added value to the pipeline. In our CNN approach, the additional information might even lead to reduced performance. The fact that the 3D network trained with 50$\times$50$\times$5 patches achieves better results than the one processing 50$\times$50$\times$11 inputs might have the same origin. Processing 2D lesion patches by a 2D Siamese network seems to be the more efficient and effective way for our PET/CT lesion tracking approach.

As a further result of our experiments, we found the added value of the 50$\times$50$\times$5 CT/Segmented CT patches amongst the two-channel patches for the 3D networks. The additional segmented CT information helps positioning the lesion within the patch, which is more relevant in the larger 3D patches and less in 2D patches.\\

Comparing the result of our AI PET/CT lesion tracking with an accuracy of 83.3\% to other CNN approaches for lesion tracking in CT scans, we reach a similar performance level, even though our cases have the additional challenge of a high number of closely neighboring bone lesions that have to be distinguished. For a lower number of neighboring lesions, Hering et al. \cite{Hering2021} achieve 80\% test accuracy during whole-body soft-tissue lesion tracking and Rafael-Palou et al. \cite{Rafael-Palou2021} re-identify 88.8\% of pulmonary nodules in follow-up scans. However, within the group of remaining lesions, we even reach a re-identification rate of 89.3\%.

Decoupling the segmentation and tracking made it possible to recognize resolved, new, remaining, fused and split lesions. Especially split lesions can cause problems in a combined approach, which runs the tracking based on follow-up lesion segmentation.\cite{Hering2021} In our pipeline, the position of the lesion contained in the segmentation is processed by the patch extraction case distinction. By being fed possible patch pairs, the Siamese network only classifies between true and false patch pairs and decides independently of the fact that the same baseline lesion might already have been assigned to other follow-up lesions. With this, several assignments or no assignment of a lesion are possible, enabling the lesion tracking to identify resolved, new, split, and fused lesions.

The fact that we used a heterogeneous dataset for the training and testing of our pipeline, including images acquired by different scanners and with even different slice thicknesses in z-direction (sec. \ref{Dataset}) shows the robustness of the approach. It is capable of handling clinical data. As our developed principle of AI lesion tracking is independent of bone lesions in PSMA-PET/CT scans, it can also be applied on other types of PET/CT scans. In fact, PSMA-PET/CT scans have only been chosen as a difficult application example due to the numerous bone lesions. 

Our findings are encouraging for the use in multimodal imaging. Nevertheless, the following limitations should be noted. In our approach, inaccuracies in the segmentation by the qPSMA software\cite{Gafita2019} are forwarded to the tracking result. Especially, in cases of disease progress and growing lesions, the segmentation software partially wrongfully draws large, fused follow-up lesions spreading in several bones instead of segmenting several uptake foci separately. In this case, the completely separated lesion segmentation and tracking causes errors. This is still a challenge for our pipeline and lowers the detection rate of fused lesion cases. The used segmentation method \cite{Gafita2019} was chosen due to its specialization on segmentation of bone lesions in PSMA-PET/CT scans. However, a new and more pipeline-optimized segmentation approach is currently in development. In addition, our test set contained a high number of resolved baseline and new follow-up lesions right beside remaining lesion pairs. Even though our pipeline handles most of the neighboring lesions well, this leads to a lower success rate for resolved and new lesion detection, which is to be improved in future works. A possibility could be to integrate a transformer network structure into our pipeline, as proposed by Tang et al.\cite{Tang2022}. The network might increase the anatomical precision for the described cases. In this context, it has be to noted that for the closely located lesions human readers reach their limits, too, concerning the differentiation of split and new, or resolved and fused lesions. For this reason, the gold standard is far from being perfect. This is demonstrated by fig. \ref{fig:ExampleCNNLesionTracking}C, where a CNN is outperforming a nuclear physician's decision, proving the ground truth dataset wrong in this case.\\

\section{Conclusion}\label{sec13}

In this paper, we successfully fill the gap of AI lesion tracking in PET/CT scans, using a pipeline of segmentation, registration, a case distinction algorithm, and a Siamese Convolutional Neural Network. The AI approach is capable of handling the high amount of lesions detected by the new technology of PSMA-PET/CT.

Different training configurations of 2D and 3D network approaches were examined, achieving an overall tracking accuracy of 83\% with an AUC=0.91 and a re-identification rate for remaining lesions of 89\%. This result is comparable to tracking approaches in other imaging modalities, even though PSMA-PET/CT scans show a relatively high number of closely located lesions.

The advantage of our project is the fully automated approach. Instead of having the user select single lesions to track, our pipeline automatically tracks all existing lesions in the scans, making it applicable for patients with a high number of lesions. With this, our approach enables a new way to process higher amounts of data for tumor response assessment. The presented approach is not limited to PSMA-PET/CT scans but is applicable for every oncological PET/CT with any other radiopharmaceutical. It could revolutionize the speed and depth of PET/CT analysis. Future clinical studies are planned to evaluate its impact.




\newpage

\backmatter
\bibliographystyle{ama}
\bibliography{sn-bibliography}

\end{document}